\documentclass[letterpaper, 10 pt, conference]{ieeeconf}  

\IEEEoverridecommandlockouts                              

\usepackage[utf8]{inputenc}
\usepackage{graphicx} 			
\usepackage{booktabs} 			
\usepackage{listings} 				
\usepackage{enumerate} 		

\usepackage{tikzscale}
\usepackage{amssymb}
\usepackage{amsmath} 
\usepackage{amsthm} 
\usepackage{etoolbox}
\usepackage{xcolor}
\usepackage{standalone}
\usepackage{tabularx}
\usepackage{adjustbox}

\usepackage{pgfplotstable}
\usepackage{siunitx}
\usepackage{colortbl}
\usepackage{mathtools}
\usepackage{hyperref}
\usepackage{ragged2e}

\newcommand{\norm}[1]{\left\|#1\right\|}

\newcommand{\bmtx}{\begin{bmatrix}}
\newcommand{\emtx}{\end{bmatrix}}
\newcommand{\bsmtx}{\left[ \begin{smallmatrix}} 
\newcommand{\esmtx}{\end{smallmatrix} \right]} 
\newcommand{\bmatarray}[1]{\left[\begin{array}{#1}}
\newcommand{\ematarray}{\end{array}\right]} 

\usepackage{tikz}
\usetikzlibrary{shadings}
\usetikzlibrary{calc}
\usetikzlibrary{intersections}
\usetikzlibrary{shapes,arrows}
\usetikzlibrary{backgrounds,fit,positioning}

\newcommand{\removeSpace}{\vspace{0pt}} 

\usepackage{pgfplots}
\usepgfplotslibrary{fillbetween}

\tikzstyle{block} = [draw, rectangle, minimum height=1cm, minimum width=1cm, node distance=1.25cm]
\tikzstyle{sum} = [draw, circle, inner sep=0.1cm, node distance=1.25cm]
\tikzstyle{dot} = [draw, circle, fill, inner sep=1pt, node distance=0.75cm]
\tikzstyle{input} = [coordinate]
\tikzstyle{output} = [coordinate, node distance=1cm]
\tikzstyle{box} = [draw=white, rectangle,minimum width=1.05cm, minimum height=.8cm,, fill=white, fill opacity=0]

\tikzset{
	andgate/.pic={
		\draw (-0.4cm,-0.3cm) arc (180:0:0.4cm) -- (0.4cm,-0.5cm) -- (-0.4cm,-0.5cm) -- cycle;
		\draw (0,0.1cm) -- (0,0.5cm);},
	deadzone/.pic={
		\draw (0,-0.4cm) -- (0,0.4cm);
		\draw (-0.4cm,0) -- (0.4cm,0);
		\draw[thick] (-0.4cm,-0.2cm) -- (-0.2cm,0) -- (0.2cm,0) -- (0.4cm,0.2cm);},
	disconnector/.pic={
		\draw (-0.5cm,0) -- (-0.2cm,0em) -- (0.2cm,-0.2cm);
		\draw (0.2cm,0) -- (0.5cm,0);
		\draw[dashed] (0,-0.5cm) -- (0,-0.1cm);
		\draw (-0.2cm,-0.5cm) -- (0.2cm,-0.5cm);},
	saturation/.pic={
		\draw (0,-0.4cm) -- (0,0.4cm);
		\draw (-0.4cm,0) -- (0.4cm,0);
		\draw[thick] (-0.4cm,-0.25cm) -- (-0.25cm,-0.25cm) -- (0.25cm,0.25cm) -- (0.4cm,0.25cm);},
	transFixRotFree/.pic={
		\node[sum](anchor){};
		\draw (anchor) -- ++(0.5cm,-0.8660cm) -- ++(-1cm,0) -- (anchor);
		\draw (anchor) ++(-0.0625,-1cm) -- ++(-0.25cm,-0.25cm);
		\draw (anchor) ++(0.1875cm,-1cm) -- ++(-0.25cm,-0.25cm);
		\draw (anchor) ++(0.4375cm,-1cm) -- ++(-0.25cm,-0.25cm);
		\draw (anchor) ++(-0.3125cm,-1cm) -- ++(-0.25cm,-0.25cm);
	}
}

\definecolor{cCtrl}{rgb}{0,0,1}								

\definecolor{cMarginRoll}{RGB}{0, 85, 149} 
\definecolor{cMarginPitch}{RGB}{0, 158, 224} 
\definecolor{cMarginYaw}{RGB}{168, 211, 255} 
\definecolor{cMarginReq}{rgb}{0,0,0}	

\definecolor{cMixedSynWeights}{rgb}{0,0,0} 	
\definecolor{cCLtrans}{RGB}{0, 85, 149} 	

\definecolor{cSimNom}{RGB}{0, 85, 149}
\colorlet{cSimUnc}{cSimNom!50}
 	
\definecolor{cSimReq}{rgb}{0,0,0} 	

\newtheorem{theorem}{Theorem}

\title{\LARGE \bf
A Robust Periodic Controller for Spacecraft Attitude Tracking
}

\author{Frederik Thiele$^{1}$, Felix Biertümpfel$^{1}$, and Harald Pfifer$^{1}$
\thanks{*This research was partially supported by the European Union under Grant No. 101153910. The responsibility for the content of this paper is with its authors.}
\thanks{$^{1}$Chair of Flight Mechanics and Control, Technische Universität Dresden, Dresden, Germany. {\tt\footnotesize$\{$frederik.thiele, felix.biertuempfel, harald.pfifer$\}$@tu-dresden.de}}%
}

\begin{document}

\maketitle
\thispagestyle{empty}
\pagestyle{empty}

\renewcommand{\arraystretch}{1.1}


\begin{abstract}
This paper presents a novel approach for robust periodic attitude control of satellites. Respecting the periodicity of the satellite dynamics in the synthesis allows to achieve constant performance and robustness requirements over the orbit. The proposed design follows a mixed sensitivity control design employing a physically motivated weighting scheme. The controller is calculated using a novel structured linear time-periodic output feedback synthesis with guaranteed optimal $\boldsymbol{L_2}$-performance. The synthesis poses a convex optimization problem and avoids grid-wise evaluations of coupling conditions inherent for classical periodic $\boldsymbol{H_\infty}$-synthesis. Moreover, the controller has a transparent and easy to implement structure. A solar power plant satellite is used to demonstrate the effectiveness of the proposed method for periodic satellite attitude control. 

\end{abstract}


\section{Introduction}
\label{sec:intro}

Satellite attitude control systems are indispensable for providing precise orientation and stability to spacecrafts - a critical aspect in many missions including communication, Earth observation and scientific exploration. Continuous technological innovation has resulted in increasingly stringent pointing requirements to ensure sensor alignment, communication connectivity and adequate power generation. Moreover, robustness against a wide spectrum of uncertainties is required to guarantee mission success. A challenging aspect is the inherently time-periodic nature of the attitude control problem. The orbital mechanics result in changing environmental disturbances and variation of the satellite's dynamics, e.g., appendages that rotate over one orbit.

A suitable approach for satellite attitude tracking is linear time-periodic (LTP) control, which directly incorporates the problem's cyclical dynamics into the control synthesis \cite{Bittanti2009}. In literature, various applications of periodic control for spacecraft exist. Particularly popular has been optimal control, see \cite{Pittelkau1993}, \cite{Lovera2002} or \cite{Wisniewski2004}. In \cite{Leomanni2020}, model predictive control for periodic systems was applied to spacecraft rendezvous maneuvers. Outside of the periodic control framework, recent advances in robust control for spacecrafts have been shown in \cite{Rodrigues2022}, \cite{Burgin2023} and \cite{Burgin2025}. Naturally, this makes the linear time-periodic $H_\infty$-framework attractive, see e.g. \cite{Colaneri2000}. Periodic $H_\infty$-synthesis requires the solution of periodic Riccati differential equations (RDEs). Numerical tools for the solution of such RDEs are readily available, e.g. in \cite{Varga2005}. A drawback of the standard $H_\infty$~output feedback synthesis is its costly computation due to a bidirectional coupling of the RDEs with a spectral radius condition.

This paper contributes a novel, mixed-sensitivity based, structured LTP controller for robust satellite attitude tracking, as presented in Section~\ref{sec:RobPer}. The approach is computationally more efficient than the standard $H_\infty$-method. The synthesis only requires the subsequent solution of two uni-directionally coupled RDEs without an algebraic condition. Moreover, the approach does not sacrifice performance and the controller is optimal in terms of the induced $L_2$-norm. The synthesis is based on the formulation of a two-block problem using a normalized coprime factorization. This idea was introduced by \cite{Glover1989} for time-invariant systems and has been extended in recent years to the parameter-varying, time-varying and time-periodic case in \cite{Theis2020}, \cite{Biertuempfel2022} and \cite{Biertuempfel2024}, respectively. 

The capabilities of the proposed control approach are demonstrated using the conceptional \emph{Abacus} satellite, which is a solar power plant in geostationary orbit. Descriptions of this satellite are readily available in literature, see e.g. \cite{Wie2001}, and are summarized in Section~\ref{sec:ssps}. The rotation of the satellite's solar assembly to maintain continuous sun-pointing orientation results in time-periodic dynamics. A robust periodic controller for attitude tracking is designed in Section~\ref{sec:ctrl}. It uses an intuitive and physically interpretable tuning process based on work of \cite{Theis2020a}. Performance and robustness of the controller are verified using time-domain simulations and frequency-domain analyses.


\section{Preliminaries}
\label{sec:prelim}


\subsection{Linear Time-Periodic Systems}
A linear time-periodic system $P$ can be described by the state-space representation
\begin{equation}\label{eqn:ltp}
	\left[\begin{array}{c}
		\dot{x}(t) \\\hline	y(t)
	\end{array}\right]	= 
	\left[\begin{array}{c|c}
		A(t) & B(t)\\\hline
		C(t) & 0
	\end{array}\right]\!
	\left[\begin{array}{c}
		x(t) \\\hline u(t) + d(t) 
	\end{array}\right], 
\end{equation}
with the state vector $x(t)\in\mathbb{R}^{n_x}$, output vector $y(t)\in\mathbb{R}^{n_y}$, input vector $u(t) \in \mathbb{R}^{n_u}$ and disturbance vector $d(t) \in \mathbb{R}^{n_u}$. The system is assumed stabilizable and detectable and the state-space matrices $A,B$ and $C$ are piecewise continuous, locally bounded matrix valued periodic functions of time with fundamental period $T$ and appropriate dimensions. For example, $A \in \mathbb{R}^{n_x \times n_x}$ with $A(t) = A(t+kT)\, \forall\,k\in~\!\!\mathbb{N}_0$. Throughout the paper, the explicit time dependency is omitted if clear from context. The plant (\ref{eqn:ltp}) is assumed strictly proper for brevity. The results of this paper can be readily extended to non-strictly proper systems at the cost of more complicated formulae. 

The performance of an LTP system can be quantified by its induced $L_2$-norm from the input $[u+d]$ to the output $y$:
\begin{equation}
	\norm{{P}}_2 := \sup_{\substack{[u+d] \in L_2, \\  [u+d] \ne 0, x(0)=0}} \frac{\norm{y}_{2}}{\norm{[u+d]}_{2}},
\end{equation}
where $\norm{\cdot}_2$ denotes the $L_2$-norm of the respective signal, i.e., $\norm{y}_2 = [\int_{0}^{\infty}y(t)^Ty(t)\mathrm{d}t]^{0.5}$. An upper bound on $\norm{P}_2$ is provided by an extension of the Bounded Real Lemma to periodic systems, given in \cite{Colaneri2000}.


\subsection{Normalized Left Coprime Factorization}

The normalized left coprime factorization of (\ref{eqn:ltp}) plays a crucial part in the derivation of the robust periodic synthesis in this paper. It is given by $P=M^{-1}N$ and said to be normalized and co-isometric, i.e., $MM^\star + NN^\star = I$, where~$^\star$ denotes the adjoint. The co-isometric property guarantees that the induced $L_2$-norm of any arbitrary stable dynamical system $H$ is preserved when interconnected with $[M\, N]$, i.e. $\norm{H}_2 \equiv \norm{H[M\, N]}_2$ \cite{Ravi1992}. The stabilizability and detectability of (\ref{eqn:ltp}) are necessary and sufficient conditions for the existence of the causal operators $M$ and $N$. For more details the reader is referred to \cite{Vuglar2016}. The normalized coprime factorization is defined by the following theorem. 

\begin{theorem}[\cite{Ravi1992}]\label{thrm:coprime}
	Let $P$ be an LTP system with a stabilizable and detectable realization given by (\ref{eqn:ltp}). Let $Z(t)\geq0$ be the unique bounded symmetric $T$-periodic solution to the Riccati equation 
	\begin{equation}\label{eqn:coprimeRDE}
		\dot{Z} = AZ + ZA^T - ZC^TCZ + BB^T
	\end{equation}
	with the boundary condition set to $Z(0)=Z(T)$. Define the packed matrix notation
	\begin{equation}\label{eqn:coprimeSS}
		N \coloneqq \left[\begin{array}{c|c}
				A+\tilde{L}C & B\\\hline
				C & 0
			\end{array}\right], \;
		M \coloneqq \left[\begin{array}{c|c}
			A+\tilde{L}C & \tilde{L}\\\hline
			C & I
		\end{array}\right], 
	\end{equation}
	with $\tilde{L} = -ZC^T$. Then $M^{-1}N$ is a normalized left coprime factorization of $P$. 
\end{theorem}\begin{proof} The proof is provided in \cite{Vuglar2016}. \end{proof}



\section{Robust Periodic Control}
\label{sec:RobPer}


\subsection{Problem Formulation}

\begin{figure}
	\centering\usetikzlibrary{positioning,plotmarks, matrix, arrows, calc, shapes}
\tikzstyle{blockdiag}	= [node distance=5mm, >=stealth', semithick]
\tikzstyle{block}			= [draw, rectangle, minimum width=1.05cm, minimum height=.8cm]
\tikzstyle{sum} 			= [draw,circle,inner sep=0pt, minimum size=6pt]
\tikzstyle{gain} 			= [draw,regular polygon, regular polygon 	sides=3,thick,minimum height=3em,minimum width=4em, rotate=30]
\tikzstyle{bguide} 		= [rectangle,minimum height=3em,minimum	width=4em]
\tikzstyle{line} 			= [thick]
\tikzstyle{branch} = [circle,inner sep=0pt,minimum size=1mm,fill=black,draw=black]
\tikzstyle{guide} = [anchor=center]
\tikzstyle{connector} 	= [draw,circle,inner sep=0pt, minimum size=0.01pt, fill=black, fill opacity=0,draw opacity=0]
\tikzstyle{textbox} = [draw=white, draw opacity=0, rectangle,minimum width=1.05cm, minimum height=.8cm]

\begin{tikzpicture}[blockdiag, auto]
	
	\node[block] (Plant) {$P$};
	\node[sum, left=of Plant, xshift=0cm] (SumP) {};
	\node[branch, left=of SumP, yshift=-0.0mm,xshift=-0mm] (BranchU) {};
	
	\node[branch, left=of BranchU, yshift=-0.0mm,xshift=0mm] (BranchXi) {};
	
	\node[block, left=of BranchXi, xshift=.4cm] (F) {$F$};
	
	\node[block, left=of F, xshift=.0cm] (Q) {$Q$};
	
	\node[connector, left = of Q, xshift = 15, yshift = -7](uFeedback){};

	\node[block, dashed,left=of BranchU, xshift=.2cm, minimum width=3.5cm, minimum height=1.5cm] (Controller) {};
	
	\node[textbox, left = of F, yshift = 0.53cm, xshift = -0.8cm](K){$K$};
	
	\node[branch, left=of Q, xshift=-.2cm] (BranchE) {};
	\node[sum, left=of BranchE, xshift=0cm] (SumE) {};
	
	%
	
	\draw[<-] (SumE) -- +(0,0.5cm)node[above]{$r$};
	\draw[->] (BranchE) -- +(0, +0.5cm)node[above]{$e$};
	\draw[->] (BranchU) -- +(0, +0.5cm)node[above]{$u$};
	\draw[<-] (SumP.north) -- +(0, +0.375cm)node[above]{$d$};
	
	\draw[->] (SumP.east) -- (Plant.west);
	\draw[->] (Q.east) -- (F.west)node[above, pos = 0.5]{$\xi$};
	\draw[-] (SumE) -- (BranchE);
	\draw[->] (BranchE) -- (Q) node[pos=0.5] {};
	\draw[-] (F) -- (BranchU) node[pos=0.5] {};
	\draw[->] (BranchU) -- ($(SumP.west)-(0mm,0mm)$);
	
	\draw[->] (Plant.east) -| +(+.3cm,-1.0cm) -| (SumE.south) node[pos=0.95,swap] {$-$};
	
	\draw[->] (BranchXi) |- +(-3.15cm,-0.6cm) |- (uFeedback);

\end{tikzpicture} 
	\caption{Unweighted four-block mixed-sensitivity problem.}
	\label{blk:mixedSynUnweightedStructured}
	\removeSpace
\end{figure}
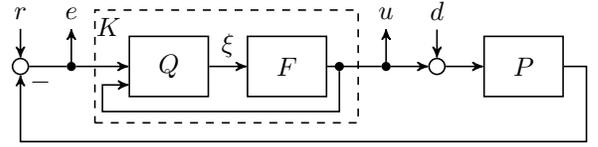

A standard feedback control problem is considered as depicted in Fig. \ref{blk:mixedSynUnweightedStructured}. For a general output feedback controller $K$, the respective closed-loop input-output map is given by
\begin{equation}
	\label{eqn:fourBlockUnweighted}
	\begin{bmatrix} e\\ u\end{bmatrix} = 
	\begin{bmatrix}S & -SP \\ KS&-KSP\end{bmatrix}
	\begin{bmatrix}r \\ d\end{bmatrix}, 
\end{equation} 
with the output sensitivity function $S=(I+PK)^{-1}$. It is referred to as the four-block mixed sensitivity problem. A standard approach in robust periodic control is the design of a controller to minimize the worst-case input-output gain, which is quantified by the induced $L_2$-norm. The general synthesis setup is suitable to pose control requirements for many engineering problems by introducing weights to the inputs and outputs \cite{Skogestad2010}. Here, an unweighted problem is chosen to expedite the derivations. The control synthesis objective is to obtain a time-periodic controller $K$ that minimizes the induced $L_2$-norm of the closed-loop interconnection of $G$ and $K$, where $G$ denotes the generalized plant depicted in Fig. \ref{blk:mixedSynUnweightedStructured}. This interconnection is given by the lower linear fractional transformation, denoted $\mathcal{F}_l(G, K)$, such that the synthesis objective can be formulated as:
\begin{equation}\label{eqn:minimization} \min_K \gamma, \textrm{s.t.} \norm{\mathcal{F}_l(G,K)}_2 \leq \gamma.
\end{equation}
Algorithms proposed in the literature, e.g. \cite{OBrien1999}, follow a two step approach to solve this output feedback problem. First, a filter RDE and a controller RDE are solved for a given $\gamma$. If both solutions exist, a spectral radius condition  is evaluated point-wise in time on a time grid in the second step. If the condition is fulfilled, a controller with closed-loop performance $\gamma$ exists and can be constructed from the system matrices and the solutions of the RDEs. A bisection over $\gamma$ provides the optimal controller. This algorithmic approach comes with notable computational overhead. 

\subsection{Structured Controller}

In contrast to classical output feedback, this paper presents a structured output feedback controller that can be synthesized by solving two unidirectionally coupled RDEs without an algebraic condition. In order to decouple the filter RDE from the controller RDE, the following structure is imposed on the controller $K$:
\begin{subequations}\label{eqn:ctrlStruct}
	\begin{align}
		\label{eqn:ctrlFilter}
		\left[\begin{array}{c}  	\dot{\xi} \\\hline \xi \end{array}\right]
		&=  
		\left[\begin{array}{c|cc} 
			A +LC &  L & B \\\hline
			I & 0  & 0  
		\end{array}\right]
		\, 
		\left[\begin{array}{c} 
			\xi \\\hline e \\ u
		\end{array}\right],
		\\
		\label{eqn:ctrlSF}
		u &= F(t)\,\xi,
	\end{align}
\end{subequations}
where the dynamics of the controller are prescribed by a filter $Q$ given by (\ref{eqn:ctrlFilter}) and a feedback gain $F$, see Fig. \ref{blk:mixedSynUnweightedStructured}. With this specific structure, properties of the normalized coprime factorization can be used to decouple the computation of $Q$ from $F$ without degrading the performance level of the overall controller (\ref{eqn:ctrlStruct}). Subsequently, a sketch of the approach is provided. The interested reader is referred to \cite{Theis2020} for a detailed derivation. 

\begin{figure}
	\centering\usetikzlibrary{positioning,plotmarks, matrix, arrows, calc, shapes}
\tikzstyle{blockdiag}	= [node distance=5mm, >=stealth', semithick]
\tikzstyle{block}			= [draw, rectangle, minimum width=1.05cm, minimum height=.8cm]
\tikzstyle{sum} 			= [draw,circle,inner sep=0pt, minimum size=6pt]
\tikzstyle{gain} 			= [draw,regular polygon, regular polygon 	sides=3,thick,minimum height=3em,minimum width=4em, rotate=30]
\tikzstyle{bguide} 		= [rectangle,minimum height=3em,minimum	width=4em]
\tikzstyle{line} 			= [thick]
\tikzstyle{branch} 		= [circle,inner sep=0pt,minimum size=1mm,fill=black,draw=black]
\tikzstyle{guide} 		= [anchor=center]

\begin{tikzpicture}[blockdiag, auto]
	

	\node[block] at (1.3,0.0)  (Minv) {$M^{-1}$};
	\draw[<-]  (Minv) -- ++(-1cm,0cm) node[pos=1,above] {$\hat{e}$};


	\node[block] at (7.6,0) (Plant) {$P$};
	\node[above=of Plant, yshift=-.3cm](Pd) {};
	\node[left=of Pd,xshift=7mm] (Wu) {};
	\node[xshift=-3mm] (SumU) at (Wu |- Plant) {};
	\node[branch, right=of SumU,xshift=-13mm] (Branch1) {};
	
	\node[left=of Branch1, xshift=3mm,minimum width=3.25cm,minimum height=1.5cm] (Controller) {};
	\node[dashed,block, xshift=0.75cm] at (Controller.center) (F) {$F$};
	\node[block, xshift=-0.75cm, yshift=0mm] at (Controller.center) (O) {$Q$};
	\node[branch, xshift=0.2cm] at (F.east) (BranchF) {};
	\node[xshift=-0.2cm, yshift=-0.25cm] at (O.west) (BranchO) {};
	
	\draw[->,dashed] ($(O.east)+(0mm,0mm)$) -- (F) node[pos=0.5] {$\xi$};
	\draw[->,dashed] (BranchF)  -- ++ (0cm,-0.6cm) -| ($(BranchO.center)-(0cm,0mm)$) -- ++ (0.2cm,0mm);
	
	\draw[->] (Controller.west) -- ($(O.west)+(0mm,0mm)$)  node[pos=0.5] {};
	\draw[-] (F) -- (BranchF.west);
	
	\node[branch, left=of Controller, xshift=4mm] (BranchE) {};
	\node[sum, left=of BranchE, xshift=3mm] (SumE) {};
	\node[above=of BranchE,yshift=0.5mm] (We) {};
	%
	
	\draw[->] (BranchE) -- +(0, +0.5cm)node[above]{$e$};
	\draw[->] (Branch1) -- +(0, +0.5cm)node[above]{$u$};
	
	\draw[->] (Minv) -- (SumE) node[pos=0.5]{};

	\draw[-] (SumE) -- (BranchE);
	\draw[-] (BranchE) -- (Controller);
	\draw[->] (BranchF) -- (Plant) node[pos=-0.15] {};
	
	\draw[->] (Plant.east)  -| +(+.3cm,-0.8cm) -| (SumE.south) node[pos=0.95,swap] {$-$};
\end{tikzpicture} 
	\caption{State feedback problem}
	\label{blk:sf_problem}
	\removeSpace
\end{figure}
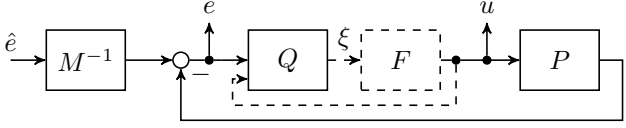

Recall that the performance of the controller is given by the induced $L_2$-norm of the closed-loop input-output map, i.e. 
\begin{equation} \label{eq:4Block}
\norm{\bmtx S & -SP \\ KS&-KSP \emtx}_2.
\end{equation}
Replacing the plant $P$ by its normalized left coprime factorization and using the co-isometric property yields
\begin{equation} \label{eq:TwoBlock}
	\norm{\bmtx S \\ KS \emtx M^{-1} \bmtx -N M\emtx}_2 = \norm{\bmtx S \\ KS \emtx M^{-1}}_2.
\end{equation}
The right hand side of (\ref{eq:TwoBlock}) represents an input-weighted two-block mixed sensitivity problem, which has the identical norm as the original four-block problem (\ref{eq:4Block}). The generalized plant of this two-block problem with the input-weight $M^{-1}$ is shown in Fig. \ref{blk:sf_problem}. By intentionally choosing the filter gain $L$ of the filter $Q$ to be equal to the solution of the normalized coprime factorization $\tilde{L}$, the two-block problem Fig. \ref{blk:sf_problem} can be transformed into a state feedback problem to obtain the controller gain $F$.
A simple state transformation yields the necessary results. The corresponding state-space representation of Fig. \ref{blk:sf_problem} can be readily derived by combining the representations of the plant (\ref{eqn:ltp}), the error filter (\ref{eqn:ctrlFilter}) and $M^{-1}$. The latter can be obtained from (\ref{eqn:coprimeSS}). After applying the linear coordinate transformation $\epsilon~=~\xi~-~(x~+~\mu)$, where $\mu$ is the internal state of $M^{-1}$, the following equivalent representation is obtained: 
\begin{equation}\label{eqn:sf_openLoop}
	\left[\begin{array}{c}
		\dot{\epsilon} \\ \dot{\mu} \\ \dot{\xi}\\ \hline e \\ u \\ \xi 
	\end{array}\right]	= 
	\left[\begin{array}{ccc|cc}
		A + L\,C& 0 & 0 & 0 & 0 \\
		0 & A & 0 & L & 0 \\
		L\,C & 0 & A & L & B \\ \hline 
		C & 0 & -C & I & 0 \\
		0 & 0 & 0 & 0 & I \\
		0 & 0 & I & 0 & 0 
	\end{array}\right]\,
	\left[\begin{array}{c}
		\epsilon\ \\ \mu \\ \xi \\ \hline \hat{e} \\ u
	\end{array}\right]. 
\end{equation}
It can be seen that $\epsilon$ is uncontrollable, $\mu$ is unobservable and the entire state $\xi$ is available, thus the gain $F$ can be determined through state feedback synthesis \cite{Theis2020}. 

The following theorem summarizes the results.
\begin{theorem}\label{thrm:structuredCtrl}
	Consider an LTP system (\ref{eqn:ltp}). There exists a controller K defined by (\ref{eqn:ctrlStruct}) such that $\norm{\mathcal{F}_l(G,K)} \leq \gamma$ if the following two conditions hold. 
	\begin{enumerate}
		\item There exists a T-periodic, continuously differentiable, symmetric positive semi-definite matrix function $Y(t)$ such that $Y$ is a stabilizing solution to 
		\begin{equation}\label{eqn:rde_filter}
			\dot{Y} = AY + YA^T - YC^T CY + B B^T.
		\end{equation}
		\item There exists a T-periodic, continuously differentiable, symmetric positive semi-definite matrix function $X(t)$ such that $X$ is a stabilizing solution to 
		\begin{equation}\label{eqn:rde_ctrl}
			\dot{X} = -\tilde{A}^TX - X\tilde{A} + X\tilde{R}X + C^T\tilde{U}C,
		\end{equation}
		
		with $\tilde{A} = A-\frac{1}{1-\gamma^2}YC^T$, $\tilde{R} = \frac{1}{1-\gamma^2}YC^TCY + B B^T$ and 
		$\tilde{U} = \frac{\gamma^2}{1-\gamma^2}C^TC$.
	\end{enumerate}
\end{theorem}
\begin{proof}
	The proof is analogous to the time-varying case in \cite{Biertuempfel2022} and requires only minor modifications.
\end{proof}

Note that the controller RDE (\ref{eqn:rde_ctrl}) and consequently the feedback gain $F = -B^TX$ still depends on $L = -YC^T$. However, the converse is not true. Hence, this novel synthesis avoids an algebraic coupling of both differential equations and, thus, the major computational drawbacks of the classical periodic $H_\infty$-synthesis, see \cite{Bittanti2009}. The decoupling process is ''loss free'' in the sense that the sequentially designed controller has the same performance level as a output feedback controller of structure (\ref{eqn:ctrlStruct}) synthesis with the standard approach.


\section{Periodic Attitude Control Problem}
\label{sec:ssps}

 
\subsection{Abacus Satellite} 
\label{sec:abacus_intro}
\begin{figure}[h!] 
	\centering\input{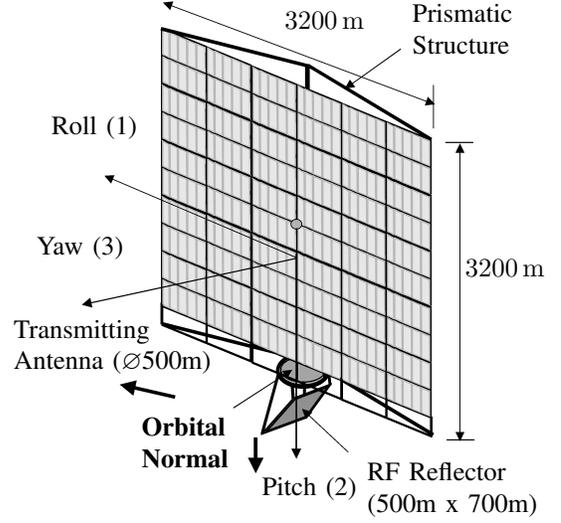}
	\caption{Schematic of \emph{Abacus} satellite \cite{Wie2001}}
	\label{fig:abacus}
	\removeSpace
\end{figure}
A periodic attitude control design for the conceptional satellite \emph{Abacus} is used to demonstrate the proposed robust synthesis. \emph{Abacus} is a solar power plant satellite in a geostationary orbit, which transmits the produced electrical energy to the surface via a microwave beam. Figure \ref{fig:abacus} shows a schematic of this satellite with its dimensions. Electrical thrusters provide three axis attitude control for sun pointing with sufficient control authority. The reader is referred to \cite{Wie2001} for a more detailed description.

Maintaining sun-pointing attitude of the solar assembly requires a continuous rotation about the satellite's pitch axis matching the orbital rate $n$. This rotation yields a periodic reference trajectory for the satellite's attitude. Reference \cite{Wie2001} provides a linear time-periodic representation of \emph{Abacus}' attitude equation of motion about this reference: 
\begin{align}\label{eq:LTPsat}
	\begin{split}
		J_1 \ddot\theta_1  + 3n^2\left(J_2 - J_3\right)[\cos^2(nt)\theta_1& + 1/2 \sin (2nt) \theta_3]\\
		&= u_1 + d_1
	\end{split}\notag\\
	\begin{split}
		J_2 \ddot\theta_2 + 3n^2\left(J_1 - J_3\right)\cos(2nt)\theta_2 &= u_2 + d_2
	\end{split}\\ 
	\begin{split}
		J_3 \ddot\theta_3  + 3n^2\left(J_2 - J_1\right)[\sin^2(nt)\theta_3&  + 1/2 \sin (2nt) \theta_1]\\
		&= u_3 + d_3,  
	\end{split}\notag
\end{align}
where $J_i$ with $i=\{1,2,3\}$ are the moments of inertia. The axis definitions can be taken from Fig.~\ref{fig:abacus}. The deviation from the reference value along the orbit are  the attitude errors $\theta_i$. The control torques and external disturbances are denoted by $u_i$ and $d_i$, respectively. Table \ref{tab:abacus} provides the numerical data for (\ref{eq:LTPsat}). The periodicity in the dynamics results from the changes of the inertia with respect to the orbital frame during the rotation in the presence of the Earth's gravitational field. Due to the satellite's orientation in the orbital frame defined in Fig. \ref{fig:abacus}, the pitch dynamics are decoupled but significantly disturbed by the gravity-gradient torque whereas the other motions are strongly coupled due to the satellite's rotation over one orbit. The reader is referred to \cite{Wie2001} for a more in-depth derivation of the equations of motion.

\begin{table}
	\caption{Properties of \emph{Abacus} satellite}
	\centering
		\vspace*{-7pt}
	\begin{tabular}{cl}
		\hline \\[-7pt]
		Orbit Rate & $n = 7.292\times10^{-5}\, \mathrm{rad}/\mathrm{sec}$ \\
		Roll Inertia & $J_1 = 2.8\times 10^{23}\, \mathrm{kg}\,\mathrm{m}^2$ \\
		Pitch Inertia & $J_2 = 1.8\times 10^{23}\, \mathrm{kg}\,\mathrm{m}^2$ \\
		Yaw Inertia & $J_3 = 4.6\times 10^{23}\, \mathrm{kg}\,\mathrm{m}^2$ \\
		\hline
	\end{tabular}
	\label{tab:abacus}
	\removeSpace
\end{table}


\subsection{External Disturbances}\label{sec:abacus_dist}
Multiple disturbance torques act on the satellite. The main environmental influences are the gravity-gradient torque and solar radiation pressure torque induced by an offset between the pressure point of the solar arrays and the satellite's center of gravity along the roll and yaw axis. The gravity-gradient torque only affects the pitch orientation due to the satellite's orientation relative to Earth and its mass distribution. The magnetic torques induced by the Earth's magnetic field can be neglected as they are compensated by appropriate routing of the electric currents in the solar assembly. An additional disturbance is caused by the microwave beam used for energy transmission. It is aligned with the pitch axis as shown in Fig. \ref{fig:abacus} and induces moments via the conservation of momentum. Due to the continuous rotation of the satellite, the microwave beam and gravity-gradient torque are cyclic disturbances over one orbit. Analytical models for the disturbances are available in \cite{Wie2001}. The worst case disturbances for the respective axes, as shown in Fig. \ref{fig:abacus} are given in newton meter as 
\begin{equation}
	\begin{aligned}
		d_1(t) &= \underbrace{2400}_{\text{solar}}-\underbrace{2380\cos(nt)}_{\text{microwave beam}} \\[3pt]
		d_2(t) &= \underbrace{240}_{\text{solar}} + \underbrace{\frac{3n^2}{10}(J_3-J_1) \sin(2nt)}_{\text{gravity-gradient}}\\[3pt]
		d_3(t) &= \underbrace{-2380\sin(nt)}_{\text{microwave beam}}.
	\end{aligned}
	\label{eqn:disturbance}
\end{equation}


\subsection{Control Objectives Definition}\label{ss:Obj}
The continuous sun-pointing of the satellite poses a periodic multi-input multi-output control problem. The satellite must track the periodic reference trajectory using the electric thrusters. The system outputs are the  attitude angles $\theta_i$ and used as feedback signals. Their absolute errors shall remain below $0.5\,$deg in all axes while rejecting the external disturbances (\ref{eqn:disturbance}). Frequency domain requirements are formulated to enforce the robustness of the closed-loop system. Specifically, the controller must satisfy one-loop-at-a-time phase and gain margins over the entire orbit of $30\,$deg and $6\,$dB, respectively. All requirements must also be satisfied for perturbations in the system dynamics resulting from $\pm20\%$ uncertainties in the satellite's moments of inertia.


\section{Periodic Control Design}
\label{sec:ctrl}


\subsection{Mixed Sensitivity Tuning Scheme}
\label{sec:design_mixedSyn}

\begin{figure}
	\centering\usetikzlibrary{positioning,plotmarks, matrix, arrows, calc, shapes}
\tikzstyle{blockdiag}	= [node distance=5mm, >=stealth', semithick]
\tikzstyle{block}			= [draw, rectangle, minimum width=1.05cm, minimum 
height=.8cm]
\tikzstyle{sum} = [draw,circle,inner sep=0pt, minimum size=6pt]
\tikzstyle{gain} = [draw,regular polygon, regular polygon 	sides=3,thick,minimum height=3em,minimum width=4em, rotate=30]
\tikzstyle{bguide} = [rectangle,minimum height=3em,minimum	width=4em]
\tikzstyle{line} = [thick]
\tikzstyle{branch} = [circle,inner sep=0pt,minimum size=1mm,fill=black,draw=black]
\tikzstyle{guide} = [anchor=center]

\begin{tikzpicture}[blockdiag, auto]
	
	\node[block, minimum height=1.0cm] (Plant) {$P$};
	\node[sum, left=of Plant, yshift=-0.0cm] (SumP) {};
	\node[branch, left=of SumP, xshift=-1.0cm] (BranchP) {};
	\node[block, above=of BranchP, xshift=0mm,yshift=0.1cm] (Wu) {$\!W_{\!u}V_{u}^{-1}\!$};
	\node[block, above=of SumP,yshift=0.05cm] (Wd) {$\!\!V_d\!\!$};
	\node[block, dashed,left=of BranchP, xshift=-0.50cm, minimum height=1.0cm] (Controller) {$K$};

	\node[above=of Plant, yshift=-.75cm, minimum width=1.2cm](Pd) {};

	\node[branch, left=of Controller, xshift=0cm] (BranchE) {};
	\node[sum, left=of BranchE, xshift=-0.72cm] (SumE) {};
	\node[block, above=of BranchE,yshift=0.10cm] (We) {$\!W_{\!e}V_{e}^{-1}\!$};
	\node[block, above=of BranchE,yshift=0.1cm, xshift=-14mm] (W1) {$\!\!V_e\!\!$};
	
	%
	%
	
	\draw[<-] (SumE) -- (W1);
	\draw[<-] (W1.north) -- +(-0,0.25cm) node[above]{$w_1$};
	\draw[<-] (Wd.north) -- +(0,0.25cm) node[above]{$w_2$};
	\draw[->] (We.north) -- +(0, +.25cm) node[above]{$z_1$};
	\draw[->] (Wu.north) -- +(0, +.25cm) node[above]{$z_2$};
	
	\draw[-] (SumE) -- (BranchE);
	\draw[->] (BranchE) -- (Controller) node[pos=0.5] {$e$};
	\draw[->] (BranchE) -- (We);
	\draw[-] (Controller) -- (BranchP) node[pos=0.3] {$u$};
	\draw[->] (Wd) -- (SumP.north);
	\draw[->] (BranchP) -- (Wu);
	\draw[->] (BranchP) -- (SumP);
	\draw[->] (SumP) -- (Plant);

	\draw[->] ($(Plant.east)+(0,-0.0cm)$) -| +(+.5cm,-0.8cm) -| (SumE.south) node[pos=0.95,swap] {$-$};
	%
	
	;\end{tikzpicture} 
	\caption{Weighted mixed sensitivity four block problem}
	\label{blk:weightedMixedSyn}
	\removeSpace
	\removeSpace
\end{figure}

The control objectives in Section \ref{ss:Obj} must be converted into meaningful closed-loop
requirements. The employed mixed-sensitivity scheme is shown in Fig.~\ref{blk:weightedMixedSyn} and the respective formulation of the generalized plant is given by 
\begin{equation}
	\begin{bmatrix}
		z_1 \\ z_2
	\end{bmatrix} \!=\! \begin{bmatrix}
		W_e V_e^{-1} &\! 0 \\ 0 &\! W_uV_u^{-1}
	\end{bmatrix}\!\!\begin{bmatrix}
		S &\! -SP \\ KS &\! -KSP
	\end{bmatrix}\!\!\begin{bmatrix}
		V_e &\! 0 \\ 0 &\! V_d
	\end{bmatrix}\!\!\begin{bmatrix}
		w_1 \\ w_2
	\end{bmatrix}.
	\label{eqn:weightedMixedSyn}
\end{equation}
In \eqref{eqn:weightedMixedSyn}, $W_u$ and $W_e$ denote dynamic weights and $V_e$, $V_u$, and $V_d$ static scaling factors. The central block is the four-block problem from \eqref{eqn:fourBlockUnweighted}. Accordingly, the corresponding weighted controller synthesis problem can be solved using the structured synthesis in Section \ref{sec:RobPer}. The proposed weights have a diagonal structure of appropriate dimension to weight physically interpretable inputs and outputs, e.g., here $W_e~=~\mathrm{diag}(W_{e,1},\;W_{e,2},\;W_{e,3})$. 

The weighting filter $W_e$ enforces the requirements on the sensitivity $S$ and disturbance sensitivity $SP$.
A high gain of $W_e$ causes a sensitivity reduction and enforces tracking and disturbance rejection capabilities.
The explicit parameterization of the weights $W_{e,i}$ with $i = \{1,2,3\}$ in the presented paper is 
\begin{equation}
	W_e(s) = \frac{s+\omega_{\mathrm{bw},i}\sqrt{\frac{3}{1-\epsilon^2}}}{2s+\epsilon\omega_{\mathrm{bw},i}\sqrt{\frac{3}{1-\epsilon^2}}}
	\label{eq:WeParam}
\end{equation}
with the constant steady state error defined by $\epsilon \ll 1$. Integral behavior up to a bandwidth $\omega_{\text{bw},i}$ and a magnitude of $0.5$ for larger frequencies is chosen. This choice reduces the sensitivity up to the desired closed-loop bandwidth and limits peak sensitivity to $2$ beyond this frequency.

The dynamic filter $W_u$ shapes the control sensitivity $KS$ and relates directly to the frequency range of the control actuation and robustness requirements. The shape of each $W_{u,i}$ is chosen with unit gain up to
a roll-off frequency $\omega_{u,i}$ and differentiating behavior afterwards.
The weights $W_{u,i}$ are parameterized as
\begin{equation}\label{eq:WuParam}
	W_{u,i}(s) = \frac{s+\omega_{u,i}}{0.01s+\omega_{u,i}}.
\end{equation}
The imposed low magnitudes at high frequencies in $KS$ above $\omega_{u,i}$ directly translate into a roll-off in the controller.

The static weights $V_e$, $V_d$ and $V_u$ act as the primary tuning knobs in the control synthesis. The ratio of $V_e$ to $V_u$ tunes the relationship between the current control error and the allowed control effort. Accordingly, the ratio between $V_d$ and $V_u$ corresponds to the control effort in response to a disturbance. A third relationship is $V_e$ to $V_d$. This ratio directly relates to the disturbance rejection performance by specifying the allowable control error for an expected disturbance level.

For the presented application, the weights are chosen time-invariant due to the constant performance and robustness requirements over the satellite's orbit. The explicit values are provided in Table \ref{tab:tuning}. The bandwidth and roll-off frequency are identical in all three channels. The former is selected to be in the order of magnitude of the orbital frequency, while the latter is chosen to provide sufficient frequency separation to the first structural mode of the satellite, which is provided in \cite{Wie2001}. The final scaling factors are obtained through an iterative tuning process. They are driven by the pointing requirement and the disturbance magnitudes of the respective axes. The significantly larger disturbance on the pitch channel is reflected in the scaling ratios.

\begin{table}
	\caption{Tuning values of controller}
	\centering
	\vspace*{-7pt}
	\begin{tabular}{cl}
			\hline \\[-7pt]
			Bandwidth & $\omega_{\mathrm{bw}} = 1.0\!\cdot\!10^{-5}\, \mathrm{rad/s}$ \\
			Roll-Off &  $\omega_{u} = 1.3\!\cdot\!10^{-3}\, \mathrm{rad/s}$  \\
			Error Scaling  & $V_e = \mathrm{diag}(0.5,\;\;0.2,\;\;0.5)$ deg \\
			Control Scaling & $V_u = \mathrm{diag}(72\;\!100,\;\;259\;\!560,\;\; 72\;\!100)$ Nm \\
			Disturbance Scaling & $V_d = \mathrm{diag}(23\;\!072,\;\; 57\;\!680,\;\;23\;\!072)$ Nm \\
			\hline
		\end{tabular}
	\label{tab:tuning}
	\removeSpace
	\removeSpace
\end{table}


\subsection{Synthesis and Implementation}

The structured robust controller is synthesized using the weights defined in Section~\ref{sec:design_mixedSyn} and
the LTP representation of the satellite in~\eqref{eq:LTPsat}. First, the gain $L$ is calculated from the solution of a scaled version of the filter RDE~\eqref{eqn:rde_filter}. This scaled RDE can be readily derived by following the step-by-step explanations in \cite{Theis2020}. The scaled RDE is solved in Matlab using a standard fourth-order Runge-Kutta fixed-step solver with a step size of $86.25\,$s. The step-size corresponds to $1000$ equidistant grid points over one orbit. The periodic filter solution is then used to determine the optimal periodic state feedback gain $F$.
This requires the iterative solution of a scaled version of the state feedback RDE~\eqref{eqn:rde_ctrl} in
a bisection over $\gamma$. The complete synthesis requires approximately $21.4\,$s on a standard laptop (Intel i7 and $16\,$GB memory) and yields an optimal performance $\gamma_\text{opt} = 1.4$. 

\begin{figure}
	\centering\input{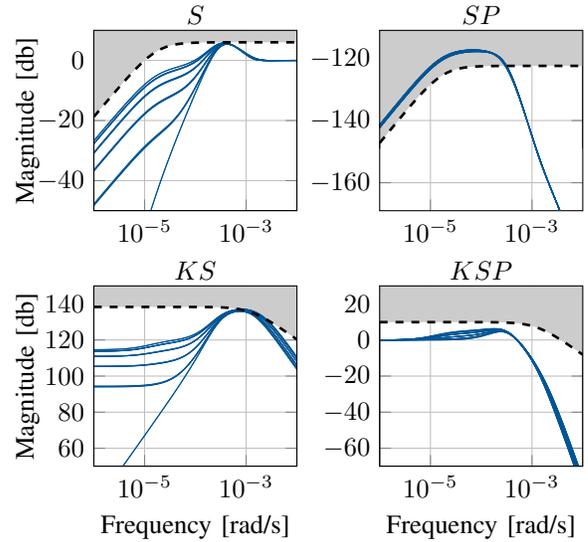}
	\caption{Comparison of achieved closed-loop transfer functions (\ref{tikz:sensitivityTransfer}) of $\theta_1$ channel with desired loop shapes (\ref{tikz:sensitivityWeight}) at twenty equidistant points over one orbit.}
	\label{fig:sensitivity}
	\removeSpace
\end{figure}

The state-space representation of the controller resulting from the weighted synthesis is
\begin{equation} 
	\left[\begin{array}{c} \dot{\xi} \\ \dot{\xi}_{W\!e} \\ \dot{\xi}_{W\!u} \\ \hline u \end{array}\right]
	= \left[\begin{array}{c|c}  		
		A_K(t)  & B_K(t) \\ \hline
		C_K(t) & D_K(t)
	\end{array}\right]
	\!
	\left[\begin{array}{c}
		\xi \\ \xi_{W\!e} \\ \xi_{W\!u} \\\hline e
	\end{array}\right] 
\end{equation}
with the system matrices 
\begin{equation*}
	\begin{aligned}
		A_K &= \begin{bmatrix}
			A+LV^{-1}_eC & 0 & 0 \\ 0 & A_{W\!e} & 0 \\ 0 & 0 & A_{W\!u}
		\end{bmatrix}+\begin{bmatrix}
			B\,F \\ 0 \\ B_{W\!u}V^{-1}_u F
		\end{bmatrix}\\[5pt]
		B_K &= [L\,V_e^{-1} \;\;\; B_{W\!e}V_e^{-1} \;\;\; 0]^T \quad C_K = F \quad D_K = 0, 
	\end{aligned}
\end{equation*}
where $\xi_{W\!e}$ and $\xi_{W\!u}$ are the internal states and $A_{W\!e}, B_{W\!e}, A_{W\!u}$ and $B_{W\!u}$ are system matrices of the weights $W_e$ and $W_u$, respectively. Note that for numerical reasons a suboptimal controller with performance index $1.1\cdot\gamma_\text{opt}$ is implemented.
Figure \ref{fig:sensitivity} depicts the corresponding closed-loop transfer functions of the $\theta_1$ channel for twenty equidistant grid points (\ref{tikz:sensitivityTransfer}) across the orbit. Comparing them to the respective weight shapes (\ref{tikz:sensitivityWeight}) verifies that the periodic controller closely follows the imposed requirements. 

\subsection{Controller Validation}
\label{sec:evaluation}

\begin{figure}
	\centering\newcommand{\plotWidth}{2.7cm}
\newcommand{\plotHeight}{2.4cm}

\newcommand{\domainStart}{400}
\newcommand{\domainEnd}{85764}

\begin{tikzpicture}
	
	\begin{axis}[%
		width= \plotWidth,
		height= \plotHeight,
		at={(0,0)},
		scale only axis,
		xmin=0,
		xmax=86164,
		xtick = {0,43082,86164,129246,172328}, 
		xticklabels = {0,0.5,1,1.5.2},
		xmajorgrids,
		ymajorgrids,
		xlabel = {Orbits},
		ymin = 5,
		ymax = 12, 
		scaled ticks=false,
		ylabel = {Gain Margin [db]},
		ylabel style = {yshift =-10},
		axis background/.style={fill=white}, 
		]
		\addplot[cMarginRoll, line width = 1] table[x expr = \thisrowno{0} ,y expr = \thisrowno{1} ,col sep=comma] {figures/tikzPlots/csv-files/GainMargin.csv};\label{tikz:marginRoll}
		\addplot[cMarginPitch, line width = 1] table[x expr = \thisrowno{0} ,y expr = \thisrowno{2} ,col sep=comma] {figures/tikzPlots/csv-files/GainMargin.csv};\label{tikz:marginPitch}
		\addplot[cMarginYaw, line width = 1] table[x expr = \thisrowno{0} ,y expr = \thisrowno{3} ,col sep=comma] {figures/tikzPlots/csv-files/GainMargin.csv};\label{tikz:marginYaw}
		\addplot[name path=gainReq, cMarginReq, dashed, line width = 1] table[row sep = crcr]{300 6\\ 85864 6\\};\label{tikz:marginReq}
		\addplot[name path=gainZero, black] table[row sep = crcr]{0 5\\86164 5\\};
		\addplot[cMarginReq!20] fill between[of = gainReq and gainZero];
		
		\addplot[black] table[row sep = crcr]{0 5\\0 13\\};
		\addplot[black] table[row sep = crcr]{86164 5\\86164 13\\};
				
	\end{axis}
	
	\begin{axis}[%
		width= \plotWidth,
		height= \plotHeight,
		at={(4.2cm,0)},
		scale only axis,
		xmin=0,
		xmax=86164,
		xmajorgrids,
		ymajorgrids,
		xtick = {0,43082,86164,129246,172328}, 
		xticklabels = {0,0.5,1,1.5.2},
		xlabel = {Orbits},
		ymin = 28,
		ymax = 36,
		scaled ticks=false,
		ylabel = {Phase Margin [deg]},
		ylabel style = {yshift =-10},
		axis background/.style={fill=white}, 
		]
		\addplot[cMarginRoll, line width = 1] table[x expr = \thisrowno{0} ,y expr = \thisrowno{1} ,col sep=comma] {figures/tikzPlots/csv-files/PhaseMargin.csv};
		\addplot[cMarginPitch, line width = 1] table[x expr = \thisrowno{0} ,y expr = \thisrowno{2} ,col sep=comma] {figures/tikzPlots/csv-files/PhaseMargin.csv};
		\addplot[cMarginYaw, line width = 1] table[x expr = \thisrowno{0} ,y expr = \thisrowno{3} ,col sep=comma] {figures/tikzPlots/csv-files/PhaseMargin.csv};
		
		\addplot[name path=phaseReq, cMarginReq, dashed, line width = 1] table[row sep = crcr]{0 30\\86164 30\\};
		\addplot[name path=phaseZero, black] table[row sep = crcr]{0 28\\86164 28\\};
		\addplot[cMarginReq!20] fill between[of = phaseReq and phaseZero];
		
		\addplot[black] table[row sep = crcr]{0 28\\0 40\\};
		\addplot[black] table[row sep = crcr]{86164 28\\86164 40\\};
	\end{axis}

\end{tikzpicture}%
	\caption{SISO margins over one orbit in roll-axis (\ref{tikz:marginRoll}), pitch-axis (\ref{tikz:marginPitch}) and yaw-axis (\ref{tikz:marginYaw}) compared with requirement \ref{tikz:marginReq}.}
	\label{fig:margin}
	\removeSpace
\end{figure}
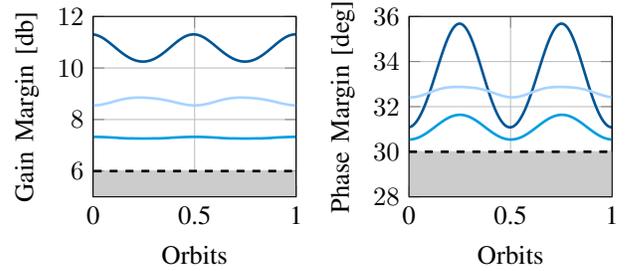

First, the robustness requirements defined in Section \ref{ss:Obj} are validated. Figure \ref{fig:margin} shows the controller's one-loop-at-a-time gain and phase margins. The gain margins remain well above $6\,$dB in all channels, with a minimum gain margin of $7.2\,$dB in the pitch channel. In general, only minor variations in the gain margins can be observed over the complete orbit, which complies with the motivation for the periodic design. The phase margins exhibit a similar behavior.
All loops satisfy the phase margin requirement of $30\,$deg, with the smallest margin being $30.5\,$deg in the pitch channel. The roll channel exhibits the largest oscillations with a $5\,$deg variation, while the other channels remain within a $1\,$deg.

\begin{figure}
	\centering\input{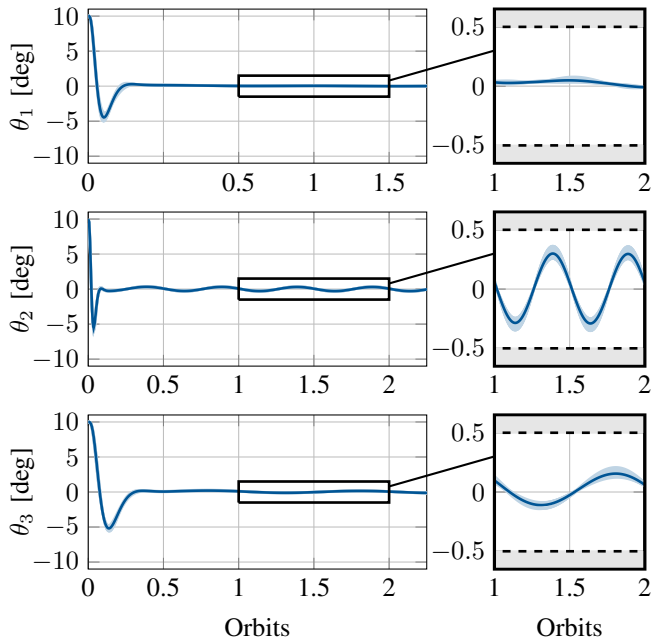}
	\caption{Simulation results with nominal inertia (\ref{tikz:resNom}) and worst-case inertia uncertainty (\ref{tikz:resUnc}) compared to pointing requirement (\ref{tikz:resReq}).}
	\label{fig:results}
	\removeSpace
	\removeSpace
\end{figure}

The controller analysis is concluded with linear time-varying simulations of the closed-loop over multiple orbits covering initial pointing errors, external disturbances, and perturbations in the satellite's dynamics. The initial pointing error is set to $10\,\mathrm{deg}\,$ in roll, pitch and yaw axis for all simulations. The attitude controller has to reject the disturbances given in (\ref{eqn:disturbance}). Moreover, a maximum of $\pm20\%$ uncertainty in the individual inertia of each axis is considered to account for model errors and fuel consumption over the satellite's lifespan. Figure \ref{fig:results} shows the simulation results. It includes the performance with nominal inertia (\ref{tikz:resNom}) as well as the worst-case permutation envelope of inertia uncertainty (\ref{tikz:resUnc}) obtained through a grid search. The controller rapidly reduces the initial pointing error and is robust for all uncertainty cases. The achieved quasi-steady state are within the requirements (\ref{tikz:resReq}), as shown in the zoomed-in region of the plot. The residual attitude oscillations can be further reduced by additional controller architectures, e.g. cyclic disturbance rejection filters as in \cite{Wie2001}, which exceeds the scope of this work. 

\section{Conclusion}
\label{sec:conclusion}
Satellite attitude control is an inherently time-periodic problem due to environmental and dynamic changes over an orbit. 
The presented control synthesis explicitly respects the satellite's periodic dynamics with a novel mixed-sensitivity synthesis approach for linear time-periodic systems. In contrast to common robust control approaches rooted in the periodic $H_\infty$-framework, the presented synthesis problem is convex and provides a norm-optimal controller. The approach's applicability for an attitude tracking problem is demonstrated on a solar power plant satellite. Future work includes the extension of the approach with internal disturbance rejection to address the remaining residual oscillation.   

%


\bibliographystyle{IEEEtran}
\bibliography{References_PeriodicSatCtrl.bib}

\begin{thebibliography}{10}
\providecommand{\url}[1]{#1}
\csname url@samestyle\endcsname
\providecommand{\newblock}{\relax}
\providecommand{\bibinfo}[2]{#2}
\providecommand{\BIBentrySTDinterwordspacing}{\spaceskip=0pt\relax}
\providecommand{\BIBentryALTinterwordstretchfactor}{4}
\providecommand{\BIBentryALTinterwordspacing}{\spaceskip=\fontdimen2\font plus
\BIBentryALTinterwordstretchfactor\fontdimen3\font minus
  \fontdimen4\font\relax}
\providecommand{\BIBforeignlanguage}[2]{{%
\expandafter\ifx\csname l@#1\endcsname\relax
\typeout{** WARNING: IEEEtran.bst: No hyphenation pattern has been}%
\typeout{** loaded for the language `#1'. Using the pattern for}%
\typeout{** the default language instead.}%
\else
\language=\csname l@#1\endcsname
\fi
#2}}
\providecommand{\BIBdecl}{\relax}
\BIBdecl

\bibitem{Bittanti2009}
S.~Bittanti and P.~Colaneri, \emph{Periodic Systems - Filtering and Control},
  ser. Communications and Control Engineering.\hskip 1em plus 0.5em minus
  0.4em\relax Springer, 2009.

\bibitem{Pittelkau1993}
M.~E. Pittelkau, ``Optimal periodic control for spacecraft pointing and
  attitude determination,'' \emph{Journal of Guidance, Control, and Dynamics},
  vol.~16, no.~6, pp. 1078--1084, 1993.

\bibitem{Lovera2002}
M.~Lovera, E.~De~Marchi, and S.~Bittanti, ``Periodic attitude control
  techniques for small satellites with magnetic actuators,'' \emph{IEEE
  Transactions on Control Systems Technology}, vol.~10, no.~1, pp. 90--95, Jan
  2002.

\bibitem{Wisniewski2004}
R.~Wisniewski and J.~Stoustrup, ``Periodic {H2} synthesis for spacecraft
  attitude control with magnetorquers,'' \emph{Journal of Guidance, Control and
  Dynamics}, vol.~27, no.~5, 2004.

\bibitem{Leomanni2020}
M.~Leomanni, G.~Bianchini, A.~Garulli, and R.~Quartullo, ``Sum-of-norms {MPC}
  for linear periodic systems with application to spacecraft rendezvous,'' in
  \emph{2020 59th IEEE Conference on Decision and Control (CDC)}, 2020, pp.
  4665--4670.

\bibitem{Rodrigues2022}
R.~Rodrigues, V.~Preda, F.~Sanfedino, and D.~Alazard, ``Modeling, robust
  control synthesis and worst-case analysis for an on-orbit servicing mission
  with large flexible spacecraft,'' \emph{Aerospace Science and Technology},
  vol. 129, p. 107865, 2022.

\bibitem{Burgin2023}
E.~Burgin, F.~Biertümpfel, and H.~Pfifer, ``Linear parameter varying
  controller design for satellite attitude control*,''
  \emph{IFAC-PapersOnLine}, vol.~56, no.~2, pp. 3112--3117, 2023, 22nd IFAC
  World Congress.

\bibitem{Burgin2025}
E.~Burgin, C.~Winkler, F.~Biert\"umpfel, H.~Pfifer, and P.~Simplicio, ``Robust
  mode transition for spacecraft attitude control,'' in \emph{AIAA SCITECH 2025
  Forum}, 2025.

\bibitem{Colaneri2000}
P.~Colaneri, ``Continuous-time periodic systems in h2 and h infinity part i:
  Theoretical aspects,'' \emph{Kybernetica}, vol.~36, pp. 211--242, 2000.

\bibitem{Varga2005}
A.~Varga, ``A periodic systems toolbox for matlab,'' \emph{IFAC Proceedings
  Volumes}, vol.~38, no.~1, pp. 450--455, 2005, 16th IFAC World Congress.

\bibitem{Glover1989}
K.~Glover and D.~McFarlane, ``Robust stabilization of normalized coprime factor
  plant descriptions with {H}-infinity-bounded uncertainty,'' \emph{IEEE
  Transactions on Automatic Control}, vol.~34, no.~8, pp. 821--830, Aug 1989.

\bibitem{Theis2020}
J.~Theis and H.~Pfifer, ``Observer‐based synthesis of linear
  parameter‐varying mixed sensitivity controllers,'' \emph{International
  Journal of Robust and Nonlinear Control}, vol.~30, no.~13, pp. 5021--5039,
  Jul. 2020.

\bibitem{Biertuempfel2022}
F.~Biertümpfel, J.~Theis, and H.~Pfifer, ``Observer-based synthesis of finite
  horizon linear time-varying controllers,'' in \emph{2022 American Control
  Conference (ACC)}, vol.~30.\hskip 1em plus 0.5em minus 0.4em\relax IEEE,
  2022.

\bibitem{Biertuempfel2024}
F.~Biertümpfel, S.~Wisbacher, H.~Pfifer, and D.~Ossmann, \emph{Periodic Robust
  Control of a Wind Turbine}, 2024.

\bibitem{Wie2001}
B.~Wie and C.~Roithmayr, ``Integrated orbit, attitude, and structural control
  system design for space solar power satellites,'' in \emph{AIAA Guidance,
  Navigation, and Control Conference and Exhibit}, 2001.

\bibitem{Theis2020a}
J.~Theis, H.~Pfifer, and P.~Seiler, ``Robust modal damping control for active
  flutter suppression,'' \emph{Journal of Guidance, Control, and Dynamics},
  vol.~43, no.~6, pp. 1056--1068, Jun. 2020.

\bibitem{Ravi1992}
R.~Ravi, A.~Pascoal, and P.~Khargonekar, ``Normalized coprime factorizations
  for linear time-varying systems,'' \emph{Systems and Control Letters},
  vol.~18, no.~6, pp. 455--465, 1992.

\bibitem{Vuglar2016}
S.~Vuglar and M.~Cantoni, ``Computing the distance between time-periodic
  dynamical systems,'' in \emph{Proceedings of 22nd International Symposium
  onMathematical Theory of Networks and Systems}, 2016.

\bibitem{Skogestad2010}
S.~Skogestad and I.~Postlethwaite, \emph{Multivariable feedback control},
  2nd~ed.\hskip 1em plus 0.5em minus 0.4em\relax Chichester: Wiley, 2010.

\bibitem{OBrien1999}
R.~O'Brien and P.~Iglesias, ``Robust controller design for linear, time-varying
  systems,'' \emph{Eur. J. Control}, vol.~5, no. 2-4, pp. 222--241, 1999.

\end{thebibliography}




\addtolength{\textheight}{-12cm}   

%

\end{document}